\documentclass[rnote]{aa}
\usepackage{graphicx}
\usepackage{txfonts}
\usepackage{natbib}

\begin{document}

\title{Recombination Line Profiles of Embedded Clusters}
\author{Sara C. Beck}

\institute{Department of Physics and Astronomy, Tel Aviv University, Ramat Aviv, Israel email: sara@wise1.tau.ac.il}

\date{Received  /Accepted}

\abstract{}{We are trying to probe conditions in the youngest super star clusters, those still 
embedded in dense obscuring clouds.}{The hydrogen recombination lines in the radio and infrared 
can be observed through the obscuration, as the optical and UV lines cannot, and give us the kinematics of the ionized gas.}{The line profiles of the clusters resemble superpositions of the lines of many very young ultra-compact or hyper-compact HII regions.  This can be explained if each 
OB star is individually embedded in dense material which it is accreting, even as it ionizes. }{We speculate on what this implies for conditions in the clusters.}


\maketitle
\section{Introduction}
Compact, populous star clusters are the main mode of star formation in many starburst galaxies.  These Super Star Clusters (SSCs) contain a few $10^5$ to  a few $10^7$ $M_\odot$ in radii 3 to 10 pc. (\citet{TB03}, \citet{J04}, \citet{Ts06}, \citet{GG07},  and references therein.) SSCs  are expected to contribute greatly to the mechanical energy, mass transport and metal enrichment of the starburst. In their mature stages, when they may be measured directly, they are observed to drive strong winds that can 
create superbubbles and strongly influence the galactic ISM \citep{Te07}. The evolution of these cluster winds is fairly well understood and the models agree reasonably well with what is observed \citep{S07}).  

But  the very youngest clusters, still deeply embedded in gas and dust clouds, are not so well understood. They cannot be observed in the optical or UV because of the extinction, but only at infrared, millimeter and radio wavelengths. The cluster
stars are not directly observable;   only the nebula they excite can be measured.  These cluster sources
are called RISN (radio-infrared supernebulae) or UDHII (ultra-dense HII regions) \citep{TB03, KJ99}.   They contain as many as 10,000 O stars in volumes no larger than a few parsecs \citep{TB04, TB03, Hea07, Be02}, yet do not have the 
cluster winds of their more evolved counterparts.  The gas motions in these clusters, as revealed in infrared and radio recombination line shapes, do not resemble O star winds.  In this note we discuss
the line shapes and what they imply for the internal structure and evolution of the nebula and cluster.   

\section{Observations}
The RISN of embedded star clusters differ from the HII regions associated with evolved SSCs. The HII regions of SSCs are usually more extensive than the stellar component .  The sizes of RISN are hard to measure. The only one which has clearly been resolved is $0.7\times0.9pc$ \citep{TB04}; for others, the size is estimated by comparing the observed radio brightness temperature $T_b$ and the electron temperature $T_e$ and is usually found to be 1-5 pc diameter.  The stars in RISN have not
been observed directly except possibly in NGC 5253 \citep{TB04} but the radio sources are so small that 
the stars cannot reasonably be confined to an even smaller voume.  So for RISN sources, the nebular gas
and the stars are coextensive. 

 The high mid-infrared fluxes of RISN argue that they are associated with a lot of hot dust.  There is dust mixed throughout the nebulae, not outside them in obscuring shells; this is demonstrated in the dependance of the measured extinction on the wavelength.  The total optical depth ($A_v$) to an RISN derived from the ratio of optical $H\alpha$ and $H\beta$ Balmer lines is typically much lower than that derived from the infrared $Br\alpha$ and $Br\gamma$ lines.  In NGC 5253 the optical lines give $A_v$ of 3 magnitudes, the infrared 18 magnitudes \citep{Tua03}. In II Zw 40 the Balmer lines give 0.2 mag $A_v$ and the Brackett lies about 8-10 magnitudes \citep{Be02}, similar to Henize 2-10 \citep{Hea07}.  The extinction derived from the ratio of thermal radio emission to $Br\alpha$ is usually even higher than that found from two Brackett lines. This means that the emitters and absorbers are mixed, so the depth to which an observation can see depends on the
wavelength of the observation, and the longer wavelengths see deeper into the source. (Note that if the measured extinction at $\lambda$, $A_\lambda$,  is greater than 1 magnitude, it implies that observations at $\lambda$ do not see through the source; this is the case for Balmer lines towards most  RISN. ) 

\subsection{Observations of Line Shapes}
The kinematics of the RISN are best studied through the infrared recombination lines $Br\alpha$ and $Br\gamma$.  These can be measured with better spatial resolution than radio recombination lines, and while not free of extinction, can usually see well into the RISN.  We have acquired $Br\alpha$ and $Br\gamma$ spectra of RISN in many nearby starbursts using the NIRSPEC spectrometer on the Keck
telescope. The spectra of He 2-10, NGC 5253 and II Zw 40 have been published \citep{Hea07, Tua03, TBC01} and M83 is in preparation \citep{CR08}.  \citet{GG07} have $Br\gamma$ spectra of many rather less embedded sources in the Antennae galaxies.  

The FWHM of the RISN lines observed ranges from less than 20  to over 100 km/sec, but the large majority fall in the range 50 to 100 km/sec. Some RISN, like that in NGC 5253, appear to have only one velocity component, and  some (those in He 2-10 for example) have a weak and broad pedestal of 
emission with FWHM  250 to 550 km/sec.  Typical RISN line shapes are shown in \citet{Tua03}, \citet{Hea07}, and \citet{CR08}. 

\section{RISN Lines and UCHII Region Lines}
The RISN lines are remarkably narrow, considering that the sources contain at least many hundreds of O stars and that O stars have strong winds at velocities of many hundreds of km/sec.  The lines do not look like O star winds, even less like
Wolf-Rayet star winds; what they resemble most are the lines seen from Galactic Ultra-Compact and Hyper-Compact HII (UCHII and HCHII) regions ( a point made by \citet{GG07} as well).  These Galactic sources are deeply embedded, excited by one or at most a few OB stars, and have infrared line widths in the same range of FWHM as do the RISN \citep{GL99}. 

We were motivated by the similarity of line shapes to simulate the profiles that would be produced by the superposition of many 
UCHII region lines.  This supposes a simple physical model in which each OB star is in its own UCHII region and the winds do not interact with each other.  We neglect stars of lower masses, which while certainly present in the forming cluster, are not strongly ionizing and are not expected to influence the recombination line profiles. In Figure 1, we show the result of superposing 30 gaussian line shapes, each with central 
velocity of a random number in the range -10 to 10 km/sec  and FWHM of a random number in the range
30 to 100 km/sec.  The spread in the allowed central velocity was chosen, since the stars in RISN cannot be observed directly, to match stellar velocity dispersions observed in visible super star clusters and globular clusters.  Since the clusters embedded in RISN have masses similar to those of  these visible 
clusters, this should be a fair assumption. The FWHM agrees with what is seen in UCHII regions. (It is not established that UCHII region lines are gaussians, but the results do not depend on shape). In Figure 2, we show the line profiles from the RISN in NGC 5253,  a source with a narrow line and no wide component. Figure 3 has a simulated profile made of  30 random lines generated as above, with the addition of 5 sources whose central velocities were chosen in the same range but whose FWHM was selected between 200 and 500 km/sec.  This models a collection of young stars, some of which have begun to develop higher velocity stellar winds; the fast component is a typical line shape for a young star which 
is obscured but not in a UCHII region.  Figure 4 shows the profiles of two RISN in He 2-10 which have 
broad pedestals in their lines.  It may be seen that the shapes of the lines in Figures 1 and 2, and 3 and 4, agree. 

\section{Implications for the Structure of the Nebula}
The line shapes generated by superposing UCHII region winds of randomly chosen central velocity and
FWHM agree very well with the RISN line shapes; those with only low FWHM winds in the sum look
like the NGC 5253 wind, and those with the higher velocity winds look like the sources in He 2-10.  A difference is that the simulated lines have asymmetries from adding a relatively small number of sources whose central velocities differ by a significant fraction of their FWHM. This effect is expected to 
decrease as more sources are included.  The observed asymmetries in the wide component of the lines 
in He 2-10 was not random but took the sense that the red wing of $Br\gamma$, but not $Br\alpha$, 
was weakened; probably an extinction effect. 

That the RISN line shapes agree with those from a simulation of superimposed UCHII regions does not prove that the RISN actually {\it is} a superposition of UCHII regions. But it strongly suggests a relation, and that  we can try to apply what we know of UCHII regions to the as-yet unobservable volume 
inside RISN.  So what can we speculate about the internal structure of the RISN from the agreement with UCHII region line shapes?  The current models for UCHII regions \citep{Ke07, HJS93, C02} agree that the winds in these sources
start while the star is still embedded in dense material and surrounded by a disk,  that material continues to accrete onto the disk (and possibly elsewhere) in the early stages of the wind, and that
this dense material suppresses and brakes the wind.  As the surrounding material  in the disk and the cloud is removed by accretion, radiation or wind pressure, the wind becomes faster and covers more solid angle.  So dense ambient material is needed to maintain the wind in its UCHII configuration: in this model each OB star in the RISN is still embedded in its own nugget of obscuring material. This local 
material, which is in addition to any large shells of dust surrounding the entire cluster,  is presumably responsible for the extinction pattern discussed above. 

If the speculation that RISN contain many UCHII regions is correct, the evolutionary state of the RISN may relate to the line shapes as follows.  in the very youngest all the O stars are deeply embedded and their winds suppressed so the net line is narrow, like that of NGC 5253. As some stars break out of the obscuration and produce faster winds the lines look like those of He 2-10, and when all the stars are freed of their cocoons, the RISN cluster will take on the appearance of a normal super star cluster and will develop a normal cluster wind. 

\section{Conclusions}
We have demonstrated that the recombination line shapes observed in RISN are like those that would be seen from a cluster of UCHII regions. Since the line shapes of UCHII regions depend  on dense obscuring material local to the star, this implies that the cluster are still embedded in, and accreting from, their individual cocoons.  We suggest that in RISN whose lines show wider  velocity  components, some of the stars have removed their  surrounding material and developed faster winds.  Eventually, all the stars will have expelled their natal cocoons and the cluster will appear as a 
normal SSC.

\section{Acknowledgements}

 We are
grateful to Drs. Mike Shull and John Bally for discussions and hospitality, and to the referee for thoughtful comments.

\begin{table}
\caption{RISN Line Shapes and Characteristics}
\label{table:1}
\centering
\begin{tabular}{cccccc}
%\tabletypesize{\footnotesize}

%\tablewidth{0pc}
%\tablecolumns{6}
%\tablehead{
\hline
Source &  O stars & $M_{tot}$&  Radius  &  FWHM  &  FWZI  \\ 
&  & $10^6 M_{\odot}$ & pc & (km s$^{-1}$ & (km s$^{-1}$ \\
\hline
%\colhead{} 	     & \colhead{}    & \colhead{$10^6 M_{\sun}$} & \colhead{pc} & \colhead{(km s$^{-1}$)} & \colhead{(km s$^{-1}$)} \\
%}
%\startdata
NGC 5253	&  1200    & $1.0$   & $0.7\times0.9$ & 76 & no broad component  \\
II Zw 40 	& 600\  & 0.1 & 0.5     &   103 & 270 \\
He2-10 1  	&  1900&1.3   & 0.4--1  & 57--78 &  220--330\\
He 2-10 2	&  1100    &  0.7  & 0.4--1 & 57--78 &22-330   \\
He 2-10 4N&  2300   &  1.5 & 0.4--1 & 57--78 &220-330   \\
He 2-10 4S	&  2400   &   1.7   &  0.4--1 & 57--78 & 220-330  \\
He 2-10 5	& 1200    &   0.8   &  0.4--1 & 57-78 & 220--330  \\
M83 	&  530 & 0.2    & -- & 78 & no broad component  \\
\hline
\end{tabular}
\end{table}

\begin{figure}
\resizebox{\hsize}{!}{\includegraphics{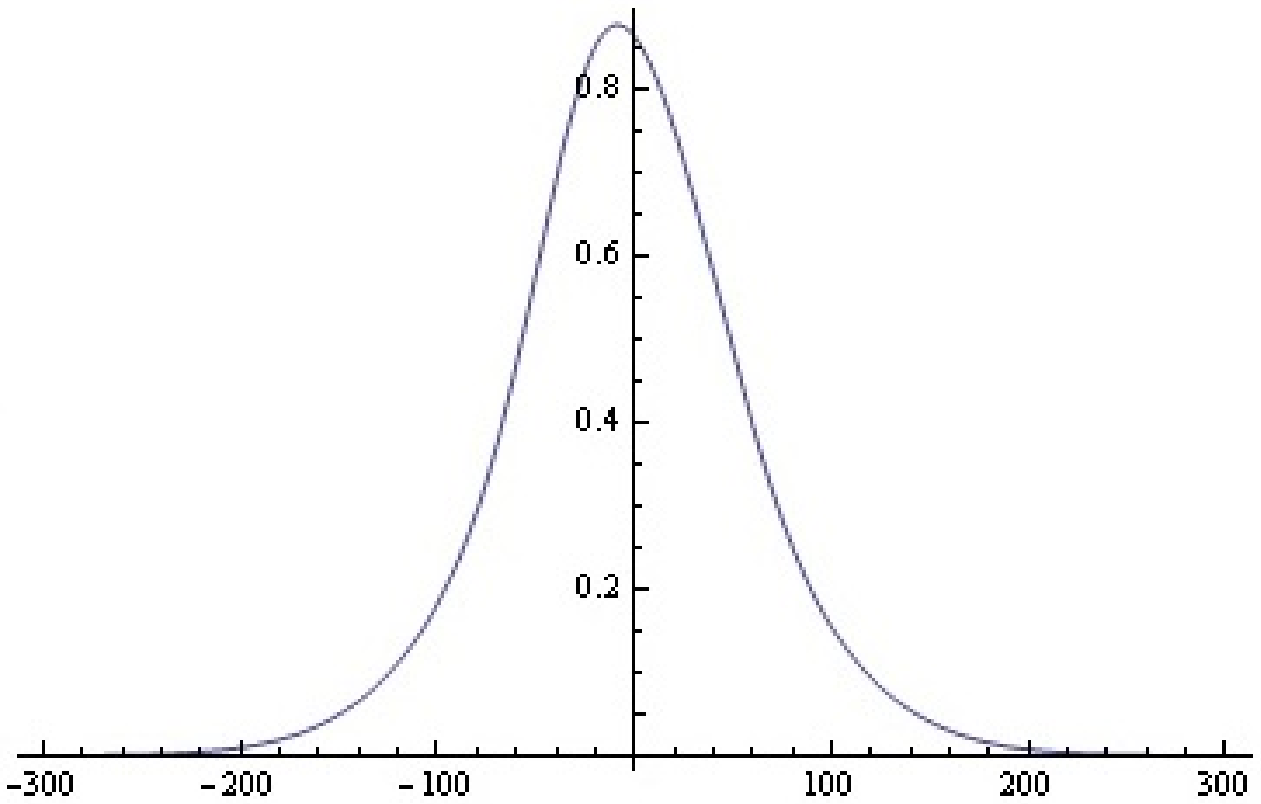}}
\caption{Simulated line shapes for a cluster of 30 UCHII regions, as described in the text.} 
\label{simulatednarrow}
\end{figure} 

\begin{figure}
\resizebox{\hsize}{!}{\includegraphics[angle=-90]{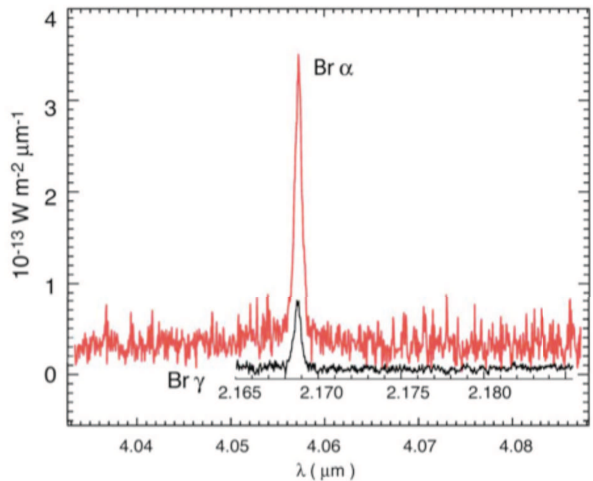}}
\caption{The Brackett line profiles of  the RISN in NGC 5253, from \citet{Tua03}.  The small ($0.001\mu$m) ticks equal 74 km/sec. This RISN has only the narrow component.}
\label{realnarrow}
\end{figure}

\begin{figure}
\resizebox{\hsize}{!}{\includegraphics{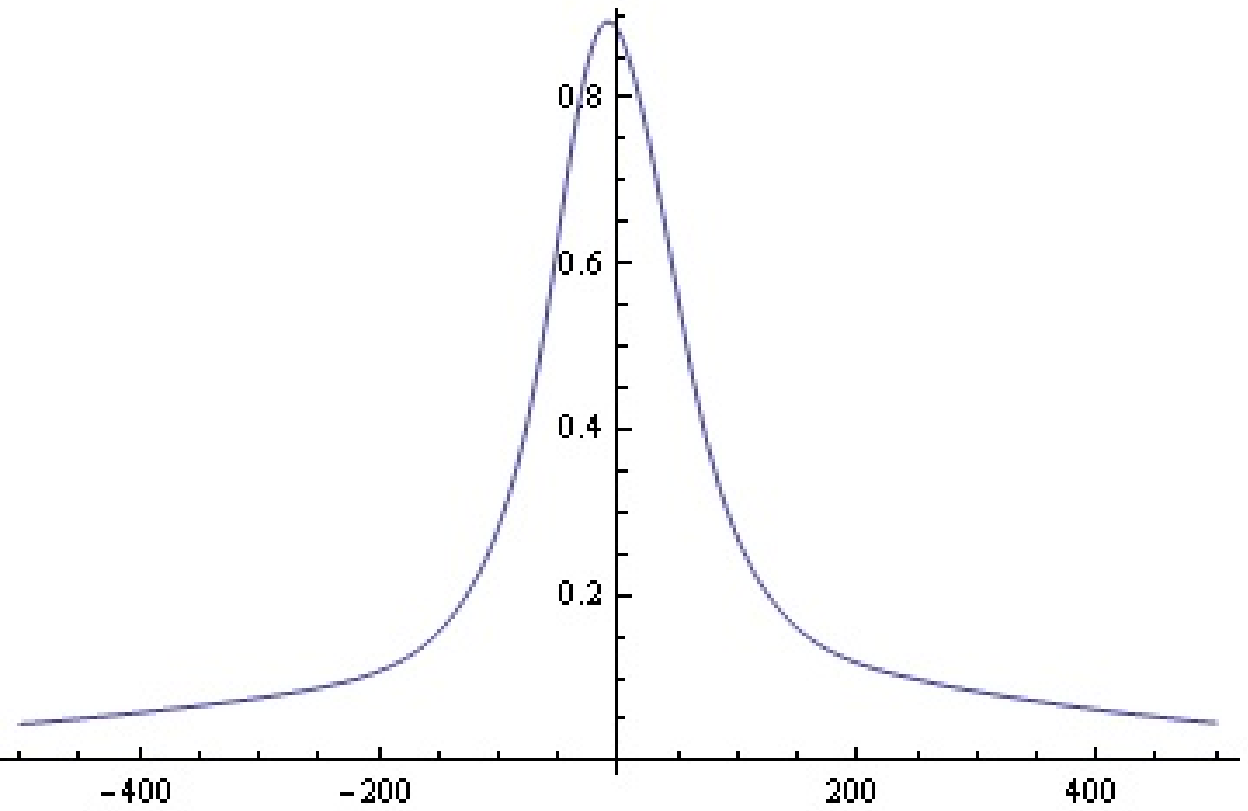}}
\caption{Simulated line shapes for a cluster of 30 UCHII regions and 5 WR stars, as described in the text} 
\label{simulatedwide}
\end{figure} 

\begin{figure}
\resizebox{\hsize}{!}{\includegraphics[angle=180]{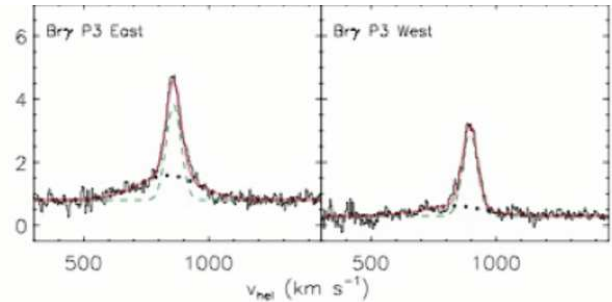}}
\caption{Brackett line profiles of two RISN in He2-10, from \citet{Hea07}.  Intensity is relative to peak. These have both 
narrow and wide components.}
\label{realwide}
\end{figure}

\end{document}